\documentclass[9pt,twocolumn,twoside]{opticajnl}

\journal{opticajournal}
\setboolean{shortarticle}{false}
\setboolean{displaycopyright}{false}

\usepackage{placeins}

\title{Broadband soliton microcombs in the biological window enabled by a resonant coupler}

\author[1,$\dagger$]{Xinrui Luo}
\author[1,$\dagger$]{Du Qian}
\author[1,2,$\dagger$]{Yuanlei Wang}
\author[1,$\dagger$]{Yiwen Yang}
\author[1]{Kaixuan Zhu}
\author[1,2]{Yinke Cheng}
\author[1]{Haoyang Luo}
\author[1]{Junqi Wang}
\author[2]{Bei-Bei Li}
\author[1,3,4]{Qihuang Gong}
\author[1,3,4,*]{Qi-Fan Yang}

\affil[1]{State Key Laboratory for Artificial Microstructure and Mesoscopic Physics and Frontiers Science Center for Nano-optoelectronics, School of Physics, Peking University, Beijing 100871, China}
\affil[2]{Beijing National Laboratory for Condensed Matter Physics, Institute of Physics, Chinese Academy of Sciences, Beijing 100190, China}
\affil[3]{Peking University Yangtze Delta Institute of Optoelectronics, Nantong, Jiangsu 226010, China}
\affil[4]{Collaborative Innovation Center of Extreme Optics, Shanxi University, Taiyuan 030006, China}
\affil[$\dagger$]{These authors contributed equally to this work.}
\affil[*]{Corresponding author: leonardoyoung@pku.edu.cn}

\begin{abstract}
Broadband soliton microcombs near 1~$\mu$m offer coherent, evenly spaced optical lines for measurements in the biological window. However, combining broad spectral coverage with closely spaced comb lines increases the pump-power requirement and makes thermally stable soliton access more difficult. Here, we demonstrate broadband silicon nitride (Si$_3$N$_4$) soliton microcombs enabled by a resonant coupler that enhances pump delivery to the nonlinear resonator. The device generates a spectrum spanning 930--1380~nm with an approximately 98-GHz comb-line spacing at on-chip pump powers of tens of milliwatts, while providing reliable thermally stable access to the single-soliton state. These results provide a route to compact, power-efficient microcomb sources that combine broad spectral coverage with closely spaced comb lines for spectroscopy and imaging in the biological window.
\end{abstract}

\dates{}
\doi{}

\begin{document}

\maketitle

\section{Introduction}

Optical frequency combs provide a precise frequency ruler for light, with many evenly spaced optical lines whose frequencies can be accurately known and controlled \cite{diddams2020optical}. Soliton microcombs bring these capabilities to integrated photonics by converting a continuous-wave laser into a broadband set of mutually coherent optical lines in a microresonator \cite{delhaye2007optical,herr2014temporal,Kippenberg2018}. Silicon nitride (Si$_3$N$_4$) has become a leading platform for this purpose because its low optical loss, broad transparency, and Kerr nonlinearity support high-performance microcombs together with large-scale photonic integration \cite{levy2010cmos,brasch2016photonic,ji2017ultra,liu2021high}. While most soliton microcombs have been developed near 1550~nm, operation near 1~$\mu$m is particularly attractive for biological imaging and spectroscopy \cite{jacques2013optical,karpov2018biological}. These applications benefit from the combination of broad optical bandwidth and closely spaced comb lines. In optical coherence tomography (OCT), the optical bandwidth sets the axial resolution, while the comb-line spacing limits the unambiguous imaging depth \cite{huang1991oct,drexler2001ultrahigh,marchand2021soliton}. In coherent Raman spectro-imaging, broad spectral coverage reaches multiple molecular vibrational bands, while closer line spacing samples their spectral structure more finely \cite{ideguchi2013coherent}.

\begin{figure*}[!t]
\centering
\includegraphics[scale=1]{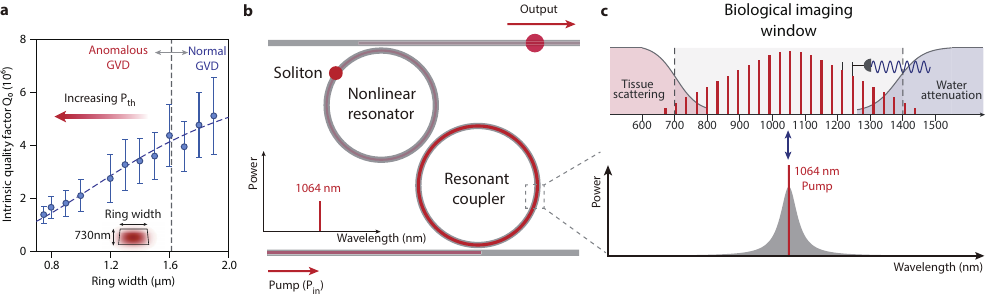}
\caption{\textbf{Dispersion--loss trade-off and resonant pumping near 1~$\mu$m.}
\textbf{a}, Measured intrinsic quality factor $Q_0$ versus ring-waveguide width. The dashed curve is a guide to the eye, and the vertical dashed line separates the anomalous- and normal-dispersion regimes at 1064~nm for the devices studied here. The horizontal arrow indicates the qualitative increase in parametric-oscillation threshold associated with lower $Q_0$. Inset: waveguide cross section and optical mode.
\textbf{b}, Resonantly pumped soliton microcomb. A continuous-wave pump at 1064~nm excites the resonant coupler (RC), which enhances pump delivery to the nonlinear resonator (NR). Solitons form in the NR, and the comb is extracted through the drop-port waveguide.
\textbf{c}, Schematic illustration of pump excitation (bottom) and broadband comb generation within the biological window (top). Tissue scattering and water attenuation indicate the short- and long-wavelength boundaries of the illustrated window.}
\label{fig:concept}
\end{figure*}

Considerable progress has been made toward broadband microcombs near 1~$\mu$m through dispersion engineering, improved fabrication, and control of the resonator spectrum \cite{karpov2018biological,yu2019tuning,briles2020generating,moille2021precursor,ou2025wafer}. However, combining broad bandwidth with closely spaced comb lines remains difficult. For the 730-nm-thick Si$_3$N$_4$ platform used here, anomalous dispersion near 1064 nm requires relatively narrow waveguides, which increases sensitivity to sidewall scattering and reduces the resonator quality factor [Fig.~\ref{fig:concept}(a)]. At the same time, reducing the repetition rate requires a larger resonator and therefore a larger mode volume. Both effects increase the pump power required for soliton generation. Consequently, broadband 1-$\mu$m soliton microcombs have typically been demonstrated with comb-line spacings approaching 1~THz \cite{karpov2018biological,yu2019tuning,briles2020generating}, whereas reducing the spacing to tens of gigahertz substantially narrows the accessible bandwidth and still requires pump powers above 100~mW \cite{lee2017towards,liu2025nearvisible}. The pump-power requirement therefore becomes a major obstacle to realizing compact, broadband microcomb systems in this wavelength range.

Here, we use resonant pumping in coupled Si$_3$N$_4$ microresonators to overcome this limitation [Fig.~\ref{fig:concept}(b)] \cite{xue2019super,zhu2026power}. A resonant coupler (RC) enhances pump delivery to a nonlinear resonator (NR), allowing broadband soliton operation without requiring a correspondingly large input pump power. Independent control of the two resonances further enables thermally stable soliton generation followed by spectral broadening at larger pump--NR detuning. With tens of milliwatts of on-chip pump power, the generated soliton microcomb spans 930--1380~nm with an approximately 98-GHz comb-line spacing [Fig.~\ref{fig:concept}(c)]. This combination of broad biological-window coverage, closely spaced comb lines, and low pump power provides a practical route toward compact microcomb sources for imaging and spectroscopy.

\section{Pump enhancement and design rules}

The main design parameters of the coupled-resonator system are the quality factors of the two resonators and their interresonator coupling rate. As established previously \cite{zhu2026power}, the RC, placed between the bus waveguide and the NR, enhances the effective pump drive of the NR. This enhancement does not arise from optical gain. Instead, it recovers the pump-coupling efficiency that would otherwise decrease rapidly at the large effective red detuning required for soliton formation. The RC provides an additional resonant pathway that coherently transfers this otherwise unused pump field to the NR.

The corresponding pump-enhancement factor is
\begin{equation}
\Gamma\approx
\frac{4G^2}
{\kappa_{\mathrm{NR}}\kappa_{\mathrm{RC}}},
\label{eq:enhancement}
\end{equation}
where $\kappa_{\mathrm{NR}}$ and $\kappa_{\mathrm{RC}}$ are the loaded angular-frequency linewidths of the NR and RC, respectively, and $G$ is the interresonator coupling rate. A large enhancement therefore requires
\begin{equation}
2G\gg\sqrt{\kappa_{\mathrm{NR}}\kappa_{\mathrm{RC}}}.
\end{equation}

The two resonators serve different functions and therefore favor different coupling conditions. The NR is the nonlinear resonator in which the soliton is formed, so a high loaded quality factor is desirable to reduce the required pump power. The RC, by contrast, is intended primarily as a linear pump-delivery resonator. Its loaded quality factor should be substantially lower to limit optical buildup and suppress unwanted nonlinear and thermal effects. Importantly, this reduction should arise predominantly from strong external coupling rather than intrinsic loss, since intrinsic loss in the RC directly dissipates pump power and reduces the efficiency of resonant power transfer. The RC should therefore retain a high intrinsic quality factor while being strongly overcoupled to the input bus waveguide. In practice, we target
\begin{equation}
Q_{\mathrm{L,RC}}\sim
\frac{1}{10}Q_{\mathrm{L,NR}}.
\label{eq:qratio}
\end{equation}
Accordingly, the RC waveguide is made sufficiently wide to mitigate intrinsic propagation loss, while its loaded linewidth is set primarily by the bus-waveguide coupling.

The interresonator coupling is then chosen sufficiently large to maintain $\Gamma\gg1$ despite the broader RC linewidth, but should not be increased arbitrarily. Excessive coupling produces stronger hybridization over the comb spectrum and also increases the pump power required for soliton initiation, even though it reduces the power required for soliton sustainment \cite{luo2026powerbandwidth}. These competing requirements therefore define the practical range of interresonator coupling.

\section{Device characterization}

\begin{figure*}[htbp]
\centering
\includegraphics[scale=1]{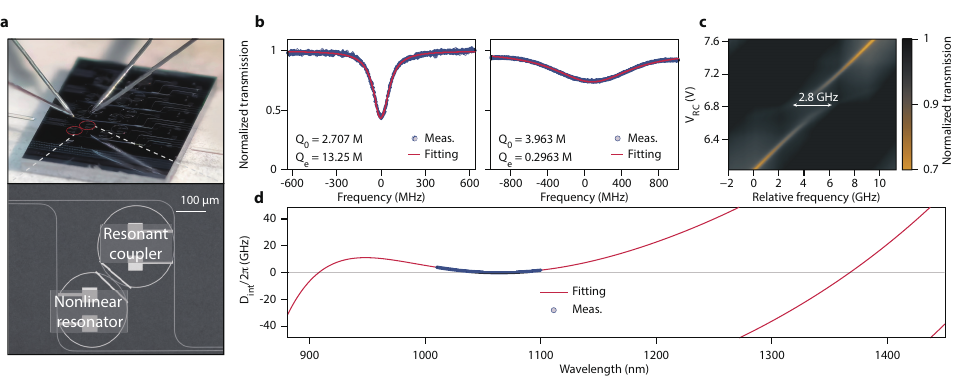}
\caption{\textbf{Passive characterization of the coupled-resonator device.}
\textbf{a}, Chip photograph and top-view micrograph of the coupled NR--RC device.
\textbf{b}, Measured resonance lineshapes of the NR and RC with corresponding fits used to extract their intrinsic and external-coupling quality factors.
\textbf{c}, Heater-controlled tuning of the RC resonance across the NR resonance, showing an avoided crossing between the two hybridized modes. The minimum mode splitting is 2.8~GHz.
\textbf{d}, Measured integrated dispersion of the NR and polynomial fit, confirming anomalous dispersion around the 1064-nm pump.}
\label{fig:passive}
\end{figure*}

Following these design rules, we fabricate the coupled-resonator device on a 730-nm-thick Si$_3$N$_4$ platform using the electron-beam-lithography-defined subtractive process reported previously \cite{wang2026compact}. The NR and RC have waveguide widths of 1.0 and 1.2~$\mu$m, respectively [Fig.~\ref{fig:passive}(a)], and both resonators are equipped with integrated heaters for independent resonance control.

We first characterize the losses and external coupling of the two resonators. Resonance fits give the intrinsic and external-coupling quality factors $(Q_0,Q_e)=(2.70,13.25)\times10^6$ for the NR and $(3.96,0.29)\times10^6$ for the RC [Fig.~\ref{fig:passive}(b)]. The corresponding loaded quality factors are $Q_{\mathrm{L,NR}}\approx2.25\times10^6$ and $Q_{\mathrm{L,RC}}\approx2.76\times10^5$. Thus, the RC retains a high intrinsic quality factor while being strongly overcoupled to the input bus waveguide, whereas the NR remains weakly coupled and maintains a high loaded quality factor. The corresponding loaded linewidths are $\kappa_{\mathrm{NR}}/2\pi\approx125$~MHz and $\kappa_{\mathrm{RC}}/2\pi\approx1.02$~GHz.

We next measure the interresonator coupling by thermally tuning an RC resonance across a fixed NR resonance. For each RC heater setting, the transmission spectrum is recorded while scanning a probe laser across the resonances. The resulting spectra reveal two hybridized resonance branches that undergo a clear avoided crossing [Fig.~\ref{fig:passive}(c)]. Fitting the extracted resonance frequencies with the eigenfrequencies of two linearly coupled resonators gives a minimum splitting of 2.8~GHz, corresponding to an interresonator coupling rate of $G/2\pi\approx1.4$~GHz \cite{haus1991coupled,Miller2015coupled}. Together with the measured linewidths, this gives a pump-enhancement factor of $\Gamma\approx61$ according to Eq.~(\ref{eq:enhancement}).

Finally, the measured integrated dispersion of the NR is shown in Fig.~\ref{fig:passive}(d). A polynomial fit gives a free spectral range $D_1/2\pi=97.72$~GHz and second-order dispersion coefficient $D_2/2\pi=388.95$~kHz, confirming anomalous dispersion around the 1064-nm pump.

\section{Thermally stable soliton generation and spectral broadening}

\begin{figure*}[!t]
\centering
\includegraphics[scale=1]{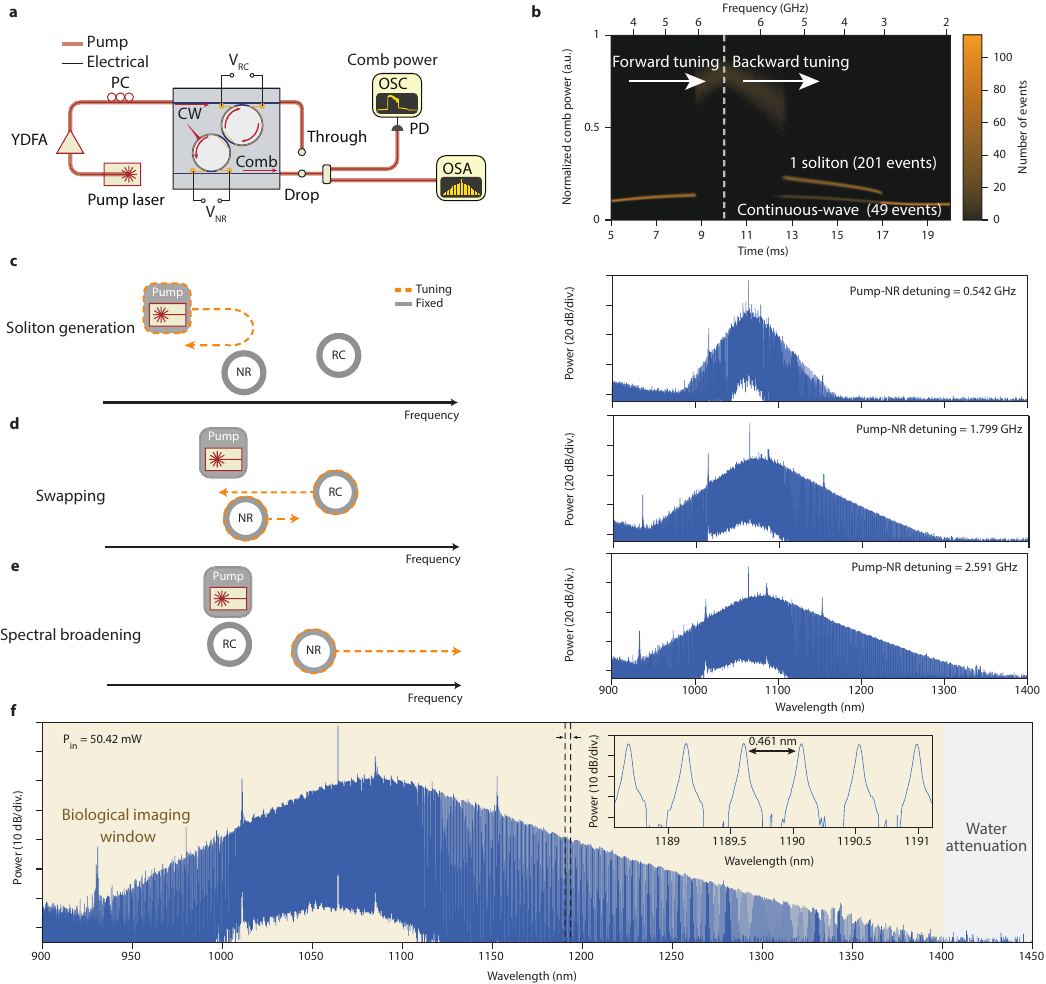}
\caption{\textbf{Thermally stable soliton generation and spectral broadening.}
\textbf{a}, Experimental setup for soliton generation and characterization. The 1064-nm pump is amplified by an ytterbium-doped fiber amplifier (YDFA) before coupling to the RC, while the comb is monitored from the NR drop port.
\textbf{b}, Comb-power trajectories from 250 repeated forward--backward tuning sequences, demonstrating reliable access to the thermally stable single-soliton branch. The single-soliton state is reached in 201 trials, while 49 trials return to the continuous-wave state.
\textbf{c--e}, Resonance-tuning sequence and corresponding soliton spectra for increasing pump--NR detuning: \textbf{c}, initial soliton state; \textbf{d}, resonance swapping; and \textbf{e}, further NR detuning for spectral broadening.
\textbf{f}, Broadband soliton spectrum spanning 930--1380~nm. The inset shows a line spacing of 0.461~nm near 1190~nm, corresponding to 97.6~GHz.
YDFA, ytterbium-doped fiber amplifier; PC, polarization controller; PD, photodetector; OSC, oscilloscope; OSA, optical spectrum analyzer; CW, continuous wave.}
\label{fig:access}
\end{figure*}

With the passive device parameters characterized, we next demonstrate thermally stable soliton generation and subsequent spectral broadening. A tunable external-cavity diode laser provides the 1064-nm continuous-wave pump, which is amplified by an ytterbium-doped fiber amplifier (YDFA) and coupled to the fundamental transverse-electric mode [Fig.~\ref{fig:access}(a)]. The through-port transmission is monitored during tuning, while the comb power and optical spectrum are measured at the NR drop port. To access the soliton state, the pump is first swept from the blue to the red side of the hybrid resonance and then swept backward while programmed heater waveforms are applied. During the backward sweep, the soliton branch is reached with the pump on the thermally stable blue side of the hybrid resonance, enabling direct soliton generation without auxiliary power modulation or active thermal stabilization \cite{zhu2026power}. Repeating the same tuning sequence 250 times yields the target single-soliton state in 201 trials, corresponding to an access probability of 80.4\% and demonstrating the reliability of this thermally stable generation scheme [Fig.~\ref{fig:access}(b)].

After soliton formation, the spectrum is broadened by progressively increasing the pump--NR detuning while maintaining resonantly enhanced pump delivery [Figs.~\ref{fig:access}(c)--\ref{fig:access}(e)]. We denote the magnitude of the pump--NR frequency detuning by $\delta_{\mathrm{NR}}$. The initial soliton spectrum corresponds to $\delta_{\mathrm{NR}}/2\pi=0.542$~GHz [Fig.~\ref{fig:access}(c)]. The RC resonance is then tuned toward the pump while the NR resonance is moved away from it. As the two bare resonances are tuned through the avoided-crossing region, their frequency ordering is exchanged while the system remains on the soliton branch [Fig.~\ref{fig:access}(d)]. This resonance-swapping process keeps the pump close to the RC resonance while increasing the pump--NR detuning, yielding $\delta_{\mathrm{NR}}/2\pi=1.799$~GHz and a broader soliton spectrum \cite{zhu2026power}.

After resonance swapping, the pump frequency and RC resonance are held fixed while the NR resonance is tuned farther from the pump. The pump--NR detuning thereby increases to $\delta_{\mathrm{NR}}/2\pi=2.591$~GHz, producing the broadest spectrum in the tuning sequence [Fig.~\ref{fig:access}(e)]. The final spectrum extends from 930 to 1380~nm above the noise floor of the optical spectrum analyzer, corresponding to approximately 105~THz of optical-frequency coverage [Fig.~\ref{fig:access}(f)]. Near 1190~nm, the measured comb-line spacing is 0.461~nm, or 97.6~GHz, consistent with the approximately 98-GHz NR free spectral range. This spacing is approximately one-tenth of the near-1-THz repetition rates of earlier broadband 1-$\mu$m soliton microcombs \cite{karpov2018biological,briles2020generating}, providing substantially denser spectral sampling across the biological window. The discrete soliton step and $\mathrm{sech}^2$-like spectral envelopes are consistent with operation on a dissipative Kerr soliton branch \cite{herr2014temporal,Kippenberg2018}.

\section{Broadband solitons at low pump power}

\begin{figure*}[htbp]
\centering
\includegraphics[scale=1]{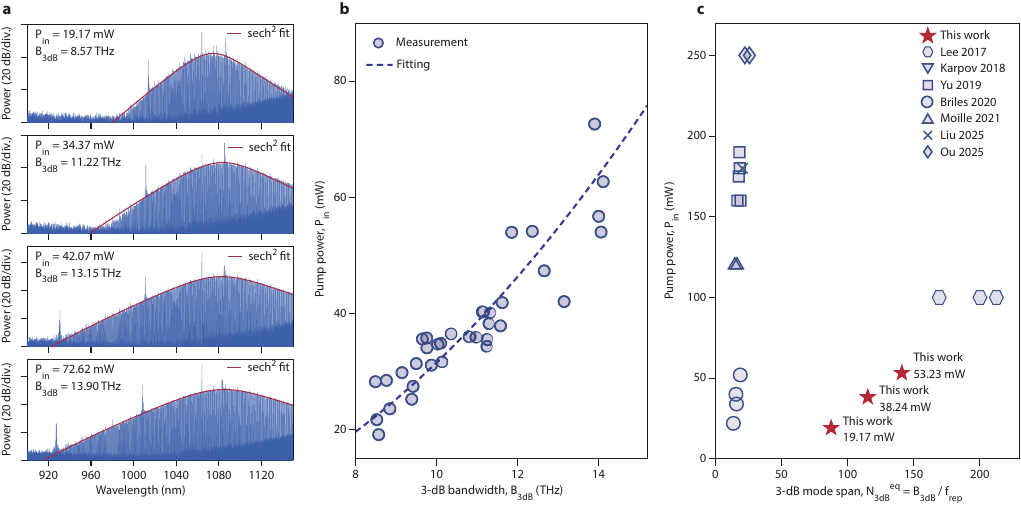}
\caption{\textbf{Broadband solitons at low pump power.}
\textbf{a}, Representative soliton spectra measured at increasing on-chip pump powers. Red curves are $\mathrm{sech}^2$ fits used to extract the full 3-dB bandwidth $B_{\mathrm{3dB}}$.
\textbf{b}, Measured on-chip pump power versus 3-dB bandwidth. Circles denote individual measurements and the dashed line is a fit to the measured trend.
\textbf{c}, Comparison with previously reported broadband soliton microcombs near 1~$\mu$m. The spectral width is expressed as the equivalent 3-dB mode span, $N_{\mathrm{3dB}}^{\mathrm{eq}}=B_{\mathrm{3dB}}/f_{\mathrm{rep}}$, to account for differences in repetition rate. Red stars denote selected measurements from this work; other symbols denote previous reports.}
\label{fig:power}
\end{figure*}

We next benchmark the pump power required for broadband soliton operation. Representative spectra are shown in Fig.~\ref{fig:power}(a). As the on-chip pump power $P_{\mathrm{in}}$ is increased from 19.17 to 72.62~mW, the fitted 3-dB bandwidth increases from 8.57 to 13.90~THz. A $\mathrm{sech}^2$ fit is applied to each measured spectrum to extract the full 3-dB bandwidth, $B_{\mathrm{3dB}}$. The complete dataset shows a clear increase of $B_{\mathrm{3dB}}$ with $P_{\mathrm{in}}$ [Fig.~\ref{fig:power}(b)].

To compare devices with different repetition rates, we further express the spectral width by the equivalent 3-dB mode span,
\begin{equation}
N_{\mathrm{3dB}}^{\mathrm{eq}}=
\frac{B_{\mathrm{3dB}}}{f_{\mathrm{rep}}},
\end{equation}
where $f_{\mathrm{rep}}$ is the comb repetition rate. This quantity approximately measures the number of comb modes contained within the 3-dB spectral bandwidth. Figure~\ref{fig:power}(c) compares the present results with previously reported broadband soliton microcombs near 1~$\mu$m \cite{lee2017towards,karpov2018biological,yu2019tuning,briles2020generating,moille2021precursor,liu2025nearvisible,ou2025wafer}. The present device occupies a distinct regime of large mode span and low pump power, reaching $N_{\mathrm{3dB}}^{\mathrm{eq}}\sim140$ with only tens of milliwatts of on-chip pump power.

\section{Discussion}

Further improvements can come from both the resonator design and the pump source. A higher intrinsic $Q$ of the NR directly lowers the nonlinear power requirement, while the RC should retain a high intrinsic $Q$ and obtain its low loaded $Q$ through strong coupling to the bus waveguide. The interresonator coupling can also be further optimized, since stronger coupling improves pump delivery but increases the power required for soliton initiation \cite{luo2026powerbandwidth}. These considerations become increasingly important as the repetition rate is reduced, where the larger resonator mode volume raises the pump requirement. Further reduction of resonator loss should make repetition rates well below 100~GHz accessible while retaining broadband operation. Another practical step is to replace the external pump-laser and amplifier chain with an integrated or co-packaged 1064-nm source. Semiconductor lasers and integrated gain near 1~$\mu$m provide possible routes toward this goal \cite{tran2022extending,zeng2026ytterbium}.

The next step is to use these sources directly in comb-based imaging and spectroscopy. Soliton microcombs have already been demonstrated as sources for spectral-domain OCT, including imaging of biological tissue \cite{marchand2021soliton}, and the broader and more densely sampled spectra targeted here could benefit this approach. Two devices with slightly different repetition rates would further provide a route to dual-comb measurements. Of particular interest are coherent Raman spectro-imaging, where dual frequency combs have enabled broadband, rapidly acquired vibrational spectra \cite{ideguchi2013coherent}, and more general dual-comb hyperspectral imaging and microscopy \cite{vicentini2021dual,mizuno2020optical}. Moving in these directions will require improved mutual coherence, control of the spectral power distribution, and integration of the comb source with the detection system.

\begin{backmatter}
\bmsection{Funding}
National Key Research and Development Program of China (2023YFB2806702); National Natural Science Foundation of China (12293050); SRAFS.

\bmsection{Disclosures}
The authors declare no conflicts of interest.

\bmsection{Data availability}
Data and simulation code supporting the findings of this study are available from the corresponding author upon reasonable request.

\bmsection{Author contributions}
X.L., D.Q., Y.W., and Y.Y. contributed equally to this work. The device was designed by X.L. and D.Q. The fabrication was performed by Y.W., Y.C., H.L., J.W., with assistance from B.-B.L. The measurement was performed by X.L., D.Q., Y.Y., and K.Z. The work was supervised by Q.G. and Q.-F.Y. 
\end{backmatter}

\bibliography{main}

\end{document}


\title{Supplementary Information: Broadband soliton microcombs in the biological window enabled by a resonant coupler}

\author{Xinrui Luo\authormark{1,$\dagger$}, Du Qian\authormark{1,$\dagger$},
Yuanlei Wang\authormark{1,2,$\dagger$}, Yiwen Yang\authormark{1,$\dagger$},
Kaixuan Zhu\authormark{1}, Yinke Cheng\authormark{1,2},
Haoyang Luo\authormark{1}, Junqi Wang\authormark{1},
Bei-Bei Li\authormark{2}, Qihuang Gong\authormark{1,3,4}, and
Qi-Fan Yang\authormark{1,3,4,*}}

\address{\authormark{1}State Key Laboratory for Artificial Microstructure and Mesoscopic Physics and Frontiers Science Center for Nano-optoelectronics, School of Physics, Peking University, Beijing 100871, China\\
\authormark{2}Beijing National Laboratory for Condensed Matter Physics, Institute of Physics, Chinese Academy of Sciences, Beijing 100190, China\\
\authormark{3}Peking University Yangtze Delta Institute of Optoelectronics, Nantong, Jiangsu 226010, China\\
\authormark{4}Collaborative Innovation Center of Extreme Optics, Shanxi University, Taiyuan 030006, China\\
\authormark{$\dagger$}These authors contributed equally to this work.}

\email{\authormark{*}leonardoyoung@pku.edu.cn}

\section{Dispersion engineering and intrinsic quality factors near 1~\texorpdfstring{$\mu$m}{um}}
\label{sec:dispersion_q}

The resonance angular frequencies of the nonlinear resonator (NR) are expanded as $\omega_\mu=\omega_0+D_1\mu+D_2\mu^2/2+\cdots$, where $\mu$ is the mode number relative to the reference resonance at $\omega_0$ and $D_1/2\pi$ is the free spectral range (FSR). The integrated dispersion is $D_{\mathrm{int}}(\mu)=\omega_\mu-(\omega_0+D_1\mu)$. Its second-order coefficient is related to the waveguide group-velocity-dispersion coefficient $\beta_2$ by $D_2=-(c/n_g)\beta_2D_1^2$, where $n_g$ is the modal group index and $c$ is the speed of light in vacuum. Anomalous dispersion therefore corresponds to $\beta_2<0$ and $D_2>0$, supporting the bright dissipative Kerr solitons considered here~\cite{herr2014temporal,Kippenberg2018}.

\begin{figure}[htbp]
\centering
\includegraphics[width=\linewidth]{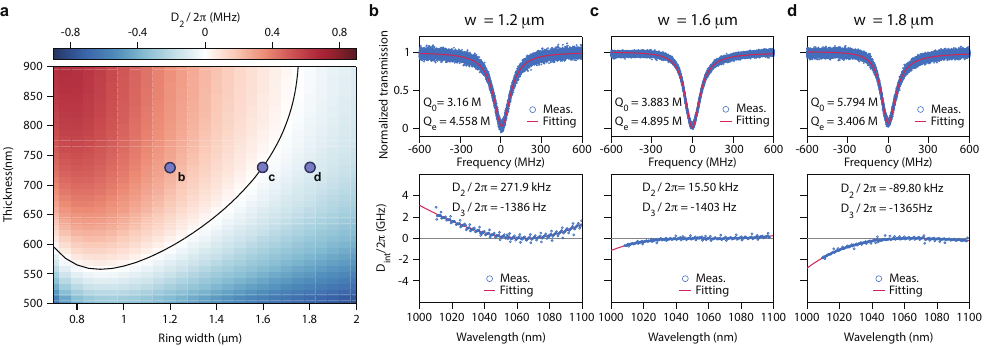}
\caption{\textbf{Dispersion engineering and intrinsic quality factors near 1~$\mu$m.}
\textbf{a}, Calculated $D_2/2\pi$ at 1064~nm for the TE$_{00}$ mode of Si$_3$N$_4$ microrings with SiO$_2$ upper and lower cladding, a radius of $R=230$~$\mu$m, and an FSR of approximately 100~GHz. The map varies the film thickness and waveguide top width. Positive values indicate anomalous dispersion, and the black curve marks $D_2=0$. At the nominal thickness of 730~nm, the zero crossing occurs near a width of 1.6~$\mu$m. Blue markers identify the geometries characterized in \textbf{b--d}.
\textbf{b--d}, Upper panels: measured resonance lineshapes and fits near 1064~nm for waveguide widths of 1.2, 1.6, and 1.8~$\mu$m, yielding intrinsic quality factors $Q_0=(3.165,3.883,5.794)\times10^6$, respectively. Lower panels: measured integrated dispersion and polynomial fits for the same devices, with $D_2/2\pi=(271.9,15.50,-89.80)$~kHz and $D_3/2\pi=(-1386,-1403,-1365)$~Hz in the same order.}
\label{fig:dispersion_q}
\end{figure}

Dispersion simulations use the fundamental transverse-electric (TE$_{00}$) mode of Si$_3$N$_4$ rings with SiO$_2$ upper and lower cladding. At the nominal film thickness of 730 nm, $D_2$ decreases with increasing waveguide width, leading to a transition from anomalous to normal dispersion ( Fig.~\ref{fig:dispersion_q}(a) ).

Measurements on representative devices follow the calculated dispersion trend. For waveguide widths of 1.2, 1.6, and 1.8~$\mu$m, polynomial fits to the measured integrated dispersion give $D_2/2\pi=(271.9,15.50,-89.80)$~kHz and $D_3/2\pi=(-1386,-1403,-1365)$~Hz, respectively (Figs.~\ref{fig:dispersion_q}(b)--\ref{fig:dispersion_q}(d)). These devices span the transition from anomalous to normal dispersion. 

Waveguide width also affects intrinsic propagation loss through the modal overlap with the etched sidewalls. Narrower waveguides are more sensitive to roughness-induced scattering~\cite{Roberts2022}. For an intrinsic power attenuation coefficient $\alpha_0$, defined by $P(z)=P(0)e^{-\alpha_0 z}$, the intrinsic quality factor is $Q_0=2\pi n_g/(\lambda_0\alpha_0)$, where $\lambda_0$ is the vacuum wavelength. Representative resonance fits near 1064~nm yield $Q_0=(3.165,3.883,5.794)\times10^6$ for widths of 1.2, 1.6, and 1.8~$\mu$m, respectively (Figs.~\ref{fig:dispersion_q}(b)--\ref{fig:dispersion_q}(d)). The increase in $Q_0$ is consistent with reduced sensitivity to sidewall scattering.

The operating NR width is selected to provide anomalous dispersion, whereas the wider resonant coupler (RC) is chosen to reduce intrinsic propagation loss. The RC loaded linewidth is set primarily by coupling to the input bus waveguide. The measured dispersion and quality factors of the operating NR are reported in Fig.~2 of the main text.

\FloatBarrier

\section{Resonant pump enhancement and soliton access}
\label{sec:effective_drive}

\subsection{Pump transfer through the resonant coupler}

We describe soliton access using a coupled Lugiato--Lefever equation (LLE) model with anomalous second-order dispersion and NR Kerr nonlinearity~\cite{zhu2026power,luo2026powerbandwidth}. The NR supports the comb-forming mode family, and the RC is represented by a linear pumped mode coupled to the pumped NR mode.

The normalized time, detunings, and coupling parameters are
\begin{equation}
\begin{gathered}
\tau=\frac{\kappa_{\mathrm{NR}}t}{2},\qquad
\zeta_j=\frac{2(\omega_{0,j}-\omega_p)}
{\kappa_{\mathrm{NR}}},
\quad j\in\{\mathrm{NR},\mathrm{RC}\},\\
K=\frac{\kappa_{\mathrm{RC}}}{\kappa_{\mathrm{NR}}},\qquad
C=\frac{2G}{\kappa_{\mathrm{NR}}},\qquad
d_2=\frac{D_2}{\kappa_{\mathrm{NR}}}.
\end{gathered}
\label{eq:normalization}
\end{equation}
Here, $t$ is the slow evolution time, $G$ is the interresonator coupling rate, and $\kappa_j=\kappa_{i,j}+\kappa_{e,j}$ is the loaded angular-frequency linewidth, comprising intrinsic and external coupling rates. The frequencies $\omega_{0,j}$ and $\omega_p$ denote the uncoupled resonances and the pump, respectively. Both detunings are normalized to the NR half-linewidth, with positive values corresponding to red detuning.

Let $A(\phi,t)$ and $B_0(t)$ denote the NR field and RC pump-mode amplitude in photon units, respectively. The single-photon Kerr frequency shift is $g_{\mathrm{NR}}=\hbar\omega_p^2cn_2/(n_0^2V_{\mathrm{eff}})$, where $n_0$, $n_2$, and $V_{\mathrm{eff}}$ denote the linear refractive index, Kerr nonlinear index, and effective NR mode volume, respectively. We then define the normalized fields as $a=\sqrt{2g_{\mathrm{NR}}/\kappa_{\mathrm{NR}}}\,A$ and $b_0=\sqrt{2g_{\mathrm{NR}}/\kappa_{\mathrm{NR}}}\,B_0$, with $a_0=\sqrt{2g_{\mathrm{NR}}/\kappa_{\mathrm{NR}}}\,A_0$ for the pumped NR mode. The normalized applied pump amplitude is $F_{\mathrm{RC}}=\sqrt{8g_{\mathrm{NR}}\kappa_{e,\mathrm{RC}}P_{\mathrm{in}}/(\hbar\omega_p\kappa_{\mathrm{NR}}^3)}$, where $P_{\mathrm{in}}$ is the on-chip pump power. The normalized coupled LLEs can then be written as\cite{zhu2026power,luo2026powerbandwidth},
\begin{align}
\frac{\partial a}{\partial\tau}
&=-(1+i\zeta_{\mathrm{NR}})a
+i d_2\frac{\partial^2a}{\partial\phi^2}
+i|a|^2a+iCb_0,\nonumber\\
\frac{db_0}{d\tau}
&=-(K+i\zeta_{\mathrm{RC}})b_0
+iCa_0+F_{\mathrm{RC}}.
\label{eq:pump_mode_equations}
\end{align}
Here, $\phi\in[0,2\pi)$ is the resonator coordinate, $a_0=(2\pi)^{-1}\int_0^{2\pi}a\,d\phi$ is the pumped NR mode, and $F_{\mathrm{RC}}$ is the normalized applied pump amplitude.

The drive acting on the NR is $F_{\mathrm{eff}}=iCb_0$, with normalized effective pump power $|F_{\mathrm{eff}}|^2=C^2|b_0|^2$. Neglecting the NR Kerr term, the linear steady state gives $a_0=iCb_0/(1+i\zeta_{\mathrm{NR}})$. Substitution into the RC equation yields
\begin{equation}
F_{\mathrm{eff}}
=\frac{iCF_{\mathrm{RC}}}
{K+i\zeta_{\mathrm{RC}}
+C^2/(1+i\zeta_{\mathrm{NR}})},
\label{eq:effective_drive}
\end{equation}
and
\begin{equation}
|F_{\mathrm{eff}}|^2=
\frac{C^2|F_{\mathrm{RC}}|^2}
{\left[K+\dfrac{C^2}{1+\zeta_{\mathrm{NR}}^2}\right]^2
+\left[\zeta_{\mathrm{RC}}
-\dfrac{C^2\zeta_{\mathrm{NR}}}
{1+\zeta_{\mathrm{NR}}^2}\right]^2}.
\label{eq:effective_drive_power}
\end{equation}
The NR contributes additional loading and a resonance shift to the RC, producing a detuning-dependent effective pump power at fixed applied pump amplitude\cite{zhu2026power,luo2026powerbandwidth}. At large NR detuning, the additional loading $C^2/(1+\zeta_{\mathrm{NR}}^2)$ and resonance shift $C^2\zeta_{\mathrm{NR}}/(1+\zeta_{\mathrm{NR}}^2)$ both approach zero. For resonant RC pumping ($\zeta_{\mathrm{RC}}=0$), Eq.~\eqref{eq:effective_drive_power} then gives $|F_{\mathrm{eff}}|^2\simeq(C^2/K^2)|F_{\mathrm{RC}}|^2$. Comparing the effective pump power with that under direct pumping of the same NR at the same input power ($8g_{\mathrm{NR}}\kappa_{e,\mathrm{NR}}P_{\mathrm{in}}/(\hbar\omega_p\kappa_{\mathrm{NR}}^3)$) gives the pump-enhancement factor~\cite{zhu2026power,luo2026powerbandwidth},
\begin{equation}
\Gamma\simeq
\frac{C^2}{K}\frac{\eta_{\mathrm{RC}}}{\eta_{\mathrm{NR}}}
=
\frac{4G^2}{\kappa_{\mathrm{NR}}\kappa_{\mathrm{RC}}}
\frac{\eta_{\mathrm{RC}}}{\eta_{\mathrm{NR}}},
\label{eq:pump_enhancement}
\end{equation}
where $\eta_j=\kappa_{e,j}/\kappa_j$ is the external coupling fraction. For the measured device, the ratio $\eta_{\mathrm{RC}}/\eta_{\mathrm{NR}}$ is of order unity. The order-of-magnitude estimate used in the main text is therefore
\begin{equation}
\Gamma\simeq
\frac{4G^2}{\kappa_{\mathrm{NR}}\kappa_{\mathrm{RC}}}.
\label{eq:enhancement_estimate}
\end{equation}

In the second-order-dispersion model, the approximate upper limit of soliton detuning is $\zeta_{\mathrm{NR}}\lesssim\pi^2|F_{\mathrm{eff}}|^2/8$~\cite{zhu2026power,luo2026powerbandwidth}. Since the soliton bandwidth scales as $B_{\mathrm{3dB}}\propto\sqrt{\zeta_{\mathrm{NR}}}$~\cite{zhu2026power}, the bandwidth near this boundary scales as $B_{\mathrm{3dB}}\propto|F_{\mathrm{eff}}|$. With the parameters of the resonant-coupler architecture held fixed, $|F_{\mathrm{eff}}|^2\propto\Gamma P_{\mathrm{in}}$ at large NR detuning and $\zeta_{\mathrm{RC}}=0$. The soliton bandwidth near the upper detuning boundary therefore scales approximately as
$B_{\mathrm{3dB}}\propto\sqrt{P_{\mathrm{in}}}$, consistent with the measured trend in Fig.~4(b) of the main text.

\begin{figure}[htbp]
\centering
\includegraphics[width=\linewidth]{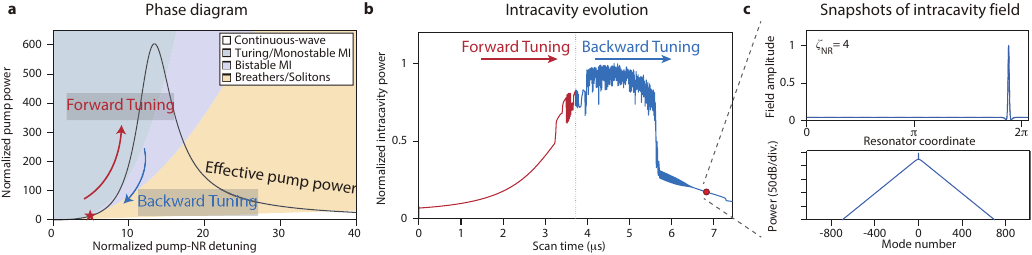}
\caption{\textbf{Forward--backward laser tuning for soliton access.}
\textbf{a}, Phase diagram of the isolated NR overlaid with the effective pump power calculated from Eqs.~\eqref{eq:effective_drive_power} and \eqref{eq:fixed_offset} (gray curve). Red and blue arrows indicate forward and backward tuning.
\textbf{b}, Simulated NR intracavity power evolution. At fixed pump amplitude and resonance offset, $\zeta_{\mathrm{NR}}$ increases linearly from $-4.0$ to $14.0$ and then decreases linearly to $2.0$. The dotted line marks the reversal, and the red circle marks the state shown in \textbf{c}.
\textbf{c}, Snapshots of the NR intracavity field (upper) and the corresponding intracavity spectrum (lower) at $\zeta_{\mathrm{NR}}\simeq4.0$.
The model parameters are $|F_{\mathrm{RC}}|=12.1$, $K=8.2$, $C=22.3$, $\Delta=25.0$, and $d_2=3.1\times10^{-3}$.}
\label{fig:phase_tuning}
\end{figure}

\subsection{Forward--backward laser tuning}

We consider laser tuning at a fixed RC--NR relative detuning $\Delta$ and neglect thermal dynamics. The two pump--resonator detunings therefore vary together according to
\begin{equation}
\zeta_{\mathrm{RC}}=\zeta_{\mathrm{NR}}+\Delta,\qquad
\Delta=\frac{2(\omega_{0,\mathrm{RC}}-\omega_{0,\mathrm{NR}})}
{\kappa_{\mathrm{NR}}}.
\label{eq:fixed_offset}
\end{equation}
Figure ~\ref{fig:phase_tuning}(a) overlays the linear steady-state response calculated from Eqs. \eqref{eq:effective_drive_power} and \eqref{eq:fixed_offset} on the phase diagram of an isolated NR under direct pumping~\cite{lugiato1987spatial,godey2014stability}. This response provides a qualitative guide to the tuning sequence, consistent with coupled-LLE simulations~\cite{zhu2026power,luo2026powerbandwidth}. During forward tuning (red arrow), increasing $\zeta_{\mathrm{NR}}$ brings the continuous-wave (CW) field into the modulation-instability (MI) region and initiates comb generation~\cite{herr2014temporal,Kippenberg2018}. Backward tuning (blue arrow) reduces the pump--NR detuning and brings the system toward the breather/soliton region, allowing a transition to the soliton state~\cite{zhu2026power}.

To illustrate this tuning sequence, we numerically solve Eq.~\eqref{eq:pump_mode_equations} using a split-step Fourier method with 2048 optical modes. Each optical mode is initially seeded with an energy corresponding to half a photon. During forward tuning, the CW field develops into a fluctuating MI comb (Fig.~\ref{fig:phase_tuning}(b)). Backward tuning produces an abrupt decrease in intracavity power, forming a soliton step. At the selected detuning, a single localized pulse and its corresponding spectral envelope are consistent with a single-soliton state (Fig.~\ref{fig:phase_tuning}(c)).

\FloatBarrier

\section{Numerical modeling of soliton states}
\label{sec:numerics}

\subsection{Generalized Lugiato--Lefever equation}

We calculate broadband soliton states using a generalized LLE that includes experimentally fitted NR dispersion and Raman effects~\cite{coen2013modeling,karpov2016raman,luo2026powerbandwidth}. The NR field is written as $A(\phi,t)=\sum_\mu A_\mu(t)e^{-i\mu\phi}$, where $|A_\mu|^2$ is the photon number in mode $\mu$. Here, $\phi=D_1t_f$ is the coordinate within one round trip, with $t_f$ denoting the fast time.

For direct pumping through the bus waveguide, the field obeys
\begin{equation}
\begin{aligned}
\frac{\partial A}{\partial t}={}&
-\left(\frac{\kappa_{\mathrm{NR}}}{2}
+i\delta_{\mathrm{NR}}\right)A
-i\widehat D_{\mathrm{int}}A
+ig_{\mathrm{NR}}|A|^2A\\
&-ig_{\mathrm{NR}}T_{\mathrm R}D_1A
\frac{\partial|A|^2}{\partial\phi}
+\sqrt{\frac{\kappa_{e,\mathrm{NR}}P_{\mathrm{in}}}
{\hbar\omega_p}}.
\end{aligned}
\label{eq:single_generalized_lle}
\end{equation}
Here, $\delta_j=\omega_{0,j}-\omega_p$ is the dimensional detuning, $T_{\mathrm R}$ is the effective Raman shock time, and $P_{\mathrm{in}}$ is the on-chip input power in the bus waveguide. The dispersion operator is
\begin{equation}
\begin{aligned}
\widehat D_{\mathrm{int}}A
&=\sum_\mu D_{\mathrm{int}}(\mu)
A_\mu e^{-i\mu\phi},\\
D_{\mathrm{int}}(\mu)
&=\sum_{j=2}^{5}\frac{D_j\mu^j}{j!},
\end{aligned}
\label{eq:dint_model}
\end{equation}
The dispersion coefficients $D_j$ are expressed in angular-frequency units, and $g_{\mathrm{NR}}$ is the single-photon Kerr frequency shift defined in Sec.~S2.1.

For resonant pumping, the bus-waveguide drive in Eq.~\eqref{eq:single_generalized_lle} is replaced by $iGB_0$, and the pumped RC mode obeys
\begin{equation}
\frac{dB_0}{dt}
=-\left(\frac{\kappa_{\mathrm{RC}}}{2}
+i\delta_{\mathrm{RC}}\right)B_0
+iGA_0
+\sqrt{\frac{\kappa_{e,\mathrm{RC}}P_{\mathrm{in}}}
{\hbar\omega_p}}.
\label{eq:coupled_rc_mode}
\end{equation}
Here, $A_0$ and $B_0$ are the mode amplitudes normalized so that their squared magnitudes give the photon numbers of the pumped NR and RC modes. The RC is treated as linear and coupled only to the pumped NR mode, consistent with full-model simulations~\cite{zhu2026power,luo2026powerbandwidth}. The fields are propagated using a split-step Fourier method with 1675 optical modes.

\subsection{Comparison of direct and resonant pumping}
\label{sec:pumping_comparison}

\begin{figure}[htbp]
\centering
\includegraphics[width=\linewidth]{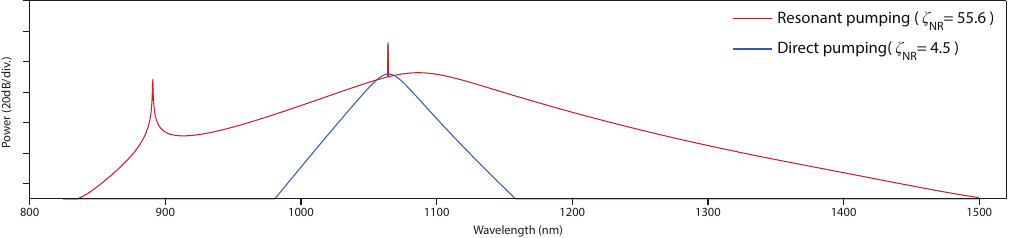}
\caption{\textbf{Soliton spectra with direct and resonant pumping.}
Simulated NR intracavity spectra at the maximum soliton detuning found for each pumping configuration, with $P_{\mathrm{in}}=39$~mW and a pump wavelength of 1064.17~nm. Resonant pumping (red) reaches $\zeta_{\mathrm{NR}}=55.6$ at $\zeta_{\mathrm{RC}}=0$, whereas direct pumping (blue) reaches $\zeta_{\mathrm{NR}}=4.5$. The corresponding 3-dB bandwidths are approximately $12.24$~THz and $4.04$~THz. Both calculations use the same NR parameters and include Raman effects with $T_{\mathrm R}=0.3$~fs. The experimentally fitted dispersion coefficients are $D_1/2\pi=97.72$~GHz, $D_2/2\pi=388.95$~kHz, $D_3/2\pi=-784.26$~Hz, $D_4/2\pi=-8.12$~Hz, and $D_5/2\pi=-0.045$~Hz. The representative quality factors used here, $(Q_{0,\mathrm{NR}},Q_{e,\mathrm{NR}})=(3.0,12.0)\times10^6$ and $(Q_{0,\mathrm{RC}},Q_{e,\mathrm{RC}})=(3.96,0.29)\times10^6$, are similar to those reported in Fig.~2 of the main text, with slight differences due to variations in $Q$ among resonances. The pump-mode splitting is $2G/2\pi=2.8$~GHz.}
\label{fig:matched_39mw_spectra}
\end{figure}

The two pumping configurations use identical NR dispersion, intrinsic loss, external coupling, Kerr coefficient, and Raman response. At $P_{\mathrm{in}}=39$~mW (a typical on-chip pump power used in the experiments), the largest pump--NR detunings found to support solitons are $\zeta_{\mathrm{NR}}=55.6$ for resonant pumping ($\zeta_{\mathrm{RC}}=0$) and $\zeta_{\mathrm{NR}}=4.5$ for direct pumping (Fig.~\ref{fig:matched_39mw_spectra}). The corresponding 3-dB bandwidths are approximately $12.24$~THz and $4.04$~THz, respectively. Resonantly enhanced pump delivery therefore supports soliton operation at larger pump–NR detuning and approximately three times the 3-dB bandwidth at the same on-chip pump power.

\bibliography{SI}